\documentclass[a4paper,11pt]{article}

\usepackage{graphicx,array}
\usepackage{color}
\usepackage{latexsym}
\usepackage{amsthm}
\usepackage{amsmath}
\usepackage{amssymb}
\usepackage{empheq}

\usepackage{dsfont}
\usepackage{epsfig}
\usepackage{slashed}
\usepackage{bbold}
\usepackage{psfrag}
\usepackage[svgnames]{xcolor}
\PassOptionsToPackage{caption=false}{subfig}
\usepackage{subcaption}
\usepackage{xfrac}
\usepackage{multirow}
\usepackage{booktabs}
\usepackage{cite}
\usepackage[normalem]{ulem}
\usepackage{jheppub} % for details on the use of the package, please see the JHEP-author-manual

\newcommand{\be}{\begin{equation}}
\newcommand{\ee}{\end{equation}}
\newcommand{\bea}{\begin{eqnarray}}
\newcommand{\eea}{\end{eqnarray}}
\newcommand{\bei}{\begin{itemize}}
\newcommand{\eei}{\end{itemize}}

\newcommand{\muc}{$\mu$C}

\newcommand{\cb}{{\cal C}_{\cal B}}
\newcommand{\cw}{{\cal C}_{\cal W}}
\newcommand{\ob}{{\cal O}_{{\cal B}}}
\newcommand{\ow}{{\cal O}_{{\cal W}}}

\title{Bottom quark electroweak dipole moments at a high-energy $\mu-$collider}

\author[a,b]{Daniele Barducci}
\affiliation[a]{Universit\`a di Pisa and INFN Section of Pisa, Largo Bruno Pontecorvo 3, 56127, Pisa, Italia}
\emailAdd{daniele.barducci@pi.infn.it}

\abstract{
We study the sensitivity of a high-energy $\mu-$collider with center of mass energy in the multi--TeV range in testing electroweak dipole interactions of the $b-$quark. We parametrize the relevant deformations in the language of the Standard Model effective field theory, where the dominant modifications arise at the $d=6$ level. 
 We analyse $\mu^+ \mu^- \to b \bar b$ and $\mu^+ \mu^- \to b \bar b h$ scatterings, performing a study at the level of a fast detector simulation. Owing the chiral structure of the dipole interaction, the study of the $\mu^+ \mu^- \to b \bar b h$ process allows to enforce the stronger bounds on the Wilson coefficients of the $d=6$ operators. The limits that can be obtained surpass present and future bounds from EW precision measurements also improving upon the ones arising from the measurement of the $\Delta F=1$ transitions $B\to X_s\gamma$.
}

\begin{document} 
Last update \today
\maketitle

%%%%%%%%%%%%%%%%%
%%% 	INTRODUCTION   %%%
%%%%%%%%%%%%%%%%%

\section{Introduction and framework}

The remarkable success of the Standard Model (SM) of particle physics has been well established by a wide range of direct searches and precision measurements, so far showing no significant discrepancies with respect to its predictions. Experimental evidences as the origin of neutrino masses, of dark matter and of the Universe baryon asymmetry as well as theoretical arguments as the flavor puzzle, the strong charge parity (CP) and Higgs hierarchy problem nevertheless strongly motivate the exploration of physics beyond the SM (BSM). 
After more than a decade of LHC operations, however, no direct sign of new physics (NP) has shown up, confirming the validity of the SM up to TeV energies. Precision tests of SM interactions can then provide a powerful avenue to probe new dynamics at energy scales much above the electroweak (EW) one.

Among the various observables, electromagnetic dipole moments of leptons have always played a special role. The magnetic moment of the electron is in fact one of the most precisely measured quantity in physics\,\cite{Fan:2022eto}
while the non observation of an associated electric dipole moment (EDM) provides strong limits on possibile CP violating effects\,\cite{Roussy:2022cmp}. Also the anomalous magnetic moment of the muon has been under intense scrutiny in the last decades, given a possible discrepancy with the theoretical predictions possibly hinting at NP effects\,\cite{Aoyama:2020ynm,Boccaletti:2024guq,Aliberti:2025beg,Muong-2:2025xyk}. Finally, recent results from the CMS collaboration have improved on previous limits on the anomalous magnetic moment of the $\tau$ lepton of around one order of magnitude\,\cite{CMS:2024qjo}, while Belle-II data effectively constrain its electric counterpart\,\cite{Belle:2021ybo}.

Moving to the dipole moments of colored fermions, they also are of extreme interest, especially for the ones of the third generation that, given their mass, might more tightly tied to the origin of the EW symmetry breaking (EWSB) mechanism. The best limits on the anomalous top quark moments are set by ATLAS and CMS by measuring the $t\bar t\gamma$ production cross-section\,\cite{CMS:2022lmh,ATLAS:2024hmk}. For what concerns the bottom quark, limits on its EDM can be derived by using neutron EDM measurements see, {\emph{e.g.}},\,\cite{Gisbert:2019ftm,Ema:2022pmo} while bounds on the anomalous magnetic moment can be set through precision EW and flavor measurements for which we refer the Reader to Sec.\,\ref{sec:current_limits}.

Looking ahead to the future, the precision measurements of the properties of 
heavy fermions is a prime target also for proposed colliders to operate in the post high luminosity LHC (HL-LHC) era.  When investigating the properties of SM particles at shorter and shorter distances, as it happens in high-energy colliders, it is however mandatory to work with the full SM symmetry restored and discuss EW dipole moments instead than the electromagnetic ones. These have been investigated at future $e^+e^-$ and $pp$ colliders for $t$, $b$ and $\tau$ see, {\emph{e.g.}},\,\cite{Rontsch:2015una,Englert:2017dev,Rajaraman:2018uyb,Durieux:2019rbz,Bissmann:2020mfi,Gu:2024wrk,Buttazzo:2026amk,Kosnik:2026cfq} while 
recent works have studied the prospects of  a future $\mu-$collider ($\mu$C) in probing $t$ and $\tau$ EW moments\,\cite{Han:2024gan,Wang:2024bfc}. On the other side a dedicated study for the case of the $b-$quark in high-energy $\mu^+\mu^-$ collisions is still lacking. In this work we fill this gap.

\subsection{Effective field theory parametrization}

We describe the SM deformations with the formalism of the  SM effective field theory (SMEFT), which allows to parametrize in a model independent way NP effects produced by heavy degrees of freedom which are present at a scale much greater of the one at which the experiment is performed.
Their effects are encoded in the Wilson coefficient $c_i^{(d)}$ of the dimension $d$ operators in the SMEFT  Lagrangian which reads
\be
{\cal L}_{\rm SMEFT} = {\cal L}_{\rm SM} + \sum_{i,\,d>4}c_i^{(d)}{\cal O}_i^{(d)} \ ,
\ee
where the mass dimension of the Wilson coefficients is $[c_i^d] = [E]^{4-d}$.
In the SMEFT the leading modifications to the down-type quark EW moments arise at $d=6$ in the momentum expansion and are parametrized by the following two operators in Warsaw basis\,\cite{Grzadkowski:2010es} 
\be\label{eq:ops}
\delta {\cal L} = \cb^{ij}\,\ob^{ij} + \cw^{ij}\,\ow^{ij} = \cb^{ij} (\bar q_L^i \sigma^{\mu\nu} d_R^j)  H B_{\mu\nu} + \cw^{ij}  (\bar q_L^i \sigma^{\mu\nu}  d_R^j) \sigma^a H W^a_{\mu\nu} \ ,
\ee
where $i,j=1,2,3$ are weak eigenstate indices, $\sigma^{\mu\nu}=\frac{i}{2}[\gamma^\mu,\gamma^\nu]$, $\sigma^a$ are the Pauli matrices, $B_{\mu\nu}$ and $W_{\mu\nu}^a$ the $U(1)_Y$ and $SU(2)_L$ field strength tensors, $q_L^i$ is left-handed quark doublet, $d_R^j$ the right-handed quark singlet, $H$ is the Higgs doublet and $\cb^{ij}$, $\cw^{ij}$ are in general complex coefficients.  

We work in the down-quark mass basis, where the $Y_d$ Yukawa matrix is diagonal and $Y_u^{\rm diag} = V Y_u$, with $V$  the CKM matrix. In this basis $q_L^i = (V_{ji}^* u_L^j,\,d_L^i)^T$, with $u_L^i$ and $d_L^i$ mass eigenstates. Since we are interested in EW dipole moments of the bottom quark, we switch on  the  $\cb^{33}$ and $\cw^{33}$ coefficients only,  rename them $\cb$ and $\cw$ hereafter and refer to the corresponding operators as $\ob$ and $\ow$. Under these assumptions and upon EWSB one obtains
\be\label{eq:ops_EWSB}
\begin{split}%
\delta{\cal L}   = & \frac{h+v}{\sqrt 2}
{\bigg{\{}} 
 (\bar b_L \sigma^{\mu\nu} b_R) 
( C_\gamma A_{\mu\nu}- {\cal C}_Z Z_{\mu\nu} ) +  (\bar u_L^k V_{kb} \sigma^{\mu\nu} b_R) \sqrt 2  \cw
W_{\mu\nu}^+ + \\
 +  & g_2  \,\cw \Big[ (\bar b_L\sigma^{\mu\nu} b_R)  W_\mu^- W_\nu^+ 
 + \sqrt 2 (\bar u_L^k V_{kb} \sigma^{\mu\nu} b_R)  (s_\omega A_\mu + c_\omega Z_\mu) W_\nu^+
 -(\mu \leftrightarrow \nu)
 {\Big{]}}   {\bigg{\}}} + h.c. \ , \\
\end{split}
\ee
with
\be
\label{eq:cgamam_cz} {\cal C}_\gamma = c_\omega\cb-s_\omega \cw \ ,  \qquad
{\cal C}_Z = s_\omega \cb+c_\omega \cw \ ,
\ee
where $k=u,c,t$, $v=246\,$GeV is the SM Higgs vacuum expectation value, $A_{\mu\nu}$, $Z_{\mu\nu}$ and $W^\pm_{\mu\nu}$ are the photon $Z-$ and $W-$boson field strength tensors and $g_2$ is the coupling constant of the $SU(2)_L$ group. In Eq.\,\eqref{eq:ops_EWSB} the first line parametrizes the  dipoles of the fermion currents with the EW bosons
 while the last line describes the terms encoding the non abelian nature of the EW interactions.  
 
 As regarding the ultraviolet (UV) origin of the effective operators $\ob$ and $\ow$, we note that the in a weakly coupled UV completion they can only arise at loop level\,\cite{Arzt:1994gp,Craig:2019wmo}, with the corresponding Wilson coefficients being rescaled by a factor $\sim 16\pi^2$ with respect to the ones of Eq.\,\eqref{eq:ops} and Eq.\,\eqref{eq:ops_EWSB}, possibly questioning the validity of the EFT description in high-energy scattering processes.
 This suppression can be lifted in strongly coupled UV scenarios see, {\emph{e.g.}},\,\cite{Giudice:2007fh,Liu:2016idz} and a related discussion in\,\cite{Han:2024gan}. We will remain agnostic regarding the UV origin of the $\ob$ and $\ow$ operators, presenting our results in terms of the Wilson coefficients of Eq.\,\eqref{eq:ops} and Eq.\,\eqref{eq:ops_EWSB} which, for concreteness, we assume to be real valued.
 
 The remainder of the paper is then organised as follows. In Sec.\,\ref{sec:current_limits} we review current limits on the dipole operators from precision EW and flavor measurements. Then, in Sec.\,\ref{sec:results} we describe the details of our computations and analyse the most relevant processes at the $\mu$C presenting our main findings. We then conclude in Sec.\,\ref{sec:concl}.

%%%%%%%%%%%%%%%%%%%%%%
%%%%% 	CURRENT LIMITS     %%%
%%%%%%%%%%%%%%%%%%%%%%

\section{Current limits on the electroweak dipole operators}\label{sec:current_limits}

The dipole operators of Eq.\,\eqref{eq:ops} are subject to various constraints arising from precision measurements of the $Z-$boson properties and neutral $\Delta F=1$ quark transitions, which we discuss in this Section in turn.
\subsection{Bounds from $R_b$}\label{sec:Rb}

Experiments at LEP and SLD precisely measured the observable $R_b = \Gamma(Z\to b\bar b)/\Gamma(Z\to q\bar q)$ for which the best fit value is $R_b^{\rm exp}\pm \delta R_b^{\rm exp} = 0.21629 \pm 0.00066$  whereas the SM prediction is $R_b^{\rm SM} = 0.21562$\,\cite{ALEPH:2005ab}. We estimate the constraint on the Wilson coefficients $\cb$ and $\cw$ from $R_b$ in the following simplified way. We analytically compute the tree-level prediction for $R_b(\cb,\cw)$ and apply a rescaling factor to match the central experimental value for $\cb=\cw=0$. 
The 95\% confidence level (CL) limits for $\cw=0$ and $\cb=0$ reads $-6.2\,{\rm TeV}^{-2} \le \cb  \le 1.1\,$TeV$^{-2}$ and 
$-3.4\,{\rm TeV}^{-2} \le \cw  \le 0.6\,$TeV$^{-2}$ respectively.
Projected limits from future experiments at CEPC\,\cite{CEPCPhysicsStudyGroup:2022uwl} and FCC-ee\,\cite{DeBlas:2019qco} will be able to decrease the relative uncertainty on $R_b$ by roughly one order of magnitude, down to $\delta R_b^{\rm exp}/R_b^{\rm exp}\simeq 2\times 10^{-4}$ and $3\times 10^{-4}$ respectively. For CEPC this results in a bound of
$-5.2\,{\rm TeV}^{-2} \le \cb  \le 0.084\,$TeV$^{-2}$ and 
$-2.9\,{\rm TeV}^{-2} \le \cw  \le 0.046\,$TeV$^{-2}$ again for $\cw=0$ and $\cb=0$ respectively.

\subsection{Bounds from $A_{\rm FB}^b$}

Another relevant constraint comes from the measurement of the forward-backward asymmetry in $e^+e^- \to b\bar b$ scattering
\be
A_{\rm FB}^b = \frac{\sigma(\cos\theta >0) - \sigma(\cos\theta<0)}{\sigma(\cos\theta >0) + \sigma(\cos\theta<0)} \ ,
\ee
where $\theta$ is the polar angle of the $b-$quark with respect to the $e^-$ direction. LEP measured $A_{\rm FB} = 0.0992\pm 0.0016$\,
  while the SM prediction is $A_{\rm FB}^{b,{\rm SM}} = 0.1037$\,\cite{ALEPH:2005ab},
 resulting in a, still unexplained, $\simeq 3\sigma$ discrepancy between theory and experiment. Here we do not aim at fitting the $A_{\rm FB}^b$ anomaly and simply provide an approximated bound on the Wilson coefficients of the operators of Eq.\,\eqref{eq:ops} with the same tree-level procedure of the one adopted for $R_b$. The extracted limits are in the same ballpark as the ones arising from $R_b$ measurements at LEP and SND. Projections for FCC-ee aims at reducing the relative error down to $10^{-3}$\,\cite{DeBlas:2019qco}. This results in a weaker limit with respect the projected one arising from $R_b$. For this observable no projections for CECP are currently available.

\subsection{Bounds from $B\to X_s\gamma$}

The $\ob$ and $\ow$ operators modify the rate for the $\Delta F = 1$ transition $B\to X_s \gamma$.
The relevant modifications are encoded by the low energy EFT (LEFT) Lagrangian
\be
{\cal L} = - \frac{4 G_F}{\sqrt 2}V_{tb} V_{ts}^* \frac{e\,m_b}{16\pi^2} 
\Bigg[
C_7 (\bar s_L \sigma^{\mu\nu} b_R) F_{\mu\nu} +
\frac{g_s}{e}  C_8   (\bar s_L \sigma^{\mu\nu} T^A b_R)  G_{\mu\nu}^A
\Bigg]  + h.c. \ ,
\ee
where $G_F$ is the Fermi constant, $e$ the electric charge, $g_s$ the strong coupling constant, $G_{\mu\nu}^A$ the gluon field strength tensor with $A=1,\,\dots,\,8$. The chromomagnetic dipole operator is included in the LEFT Lagrangian because of the large mixing effect with the electromagnetic dipole. In the down-quark mass basis $\cb$ and $\cw$ give no direct tree-level contributions to $C_7$ and $C_8$. For the one-loop contributions we use the results of \,\cite{Dekens:2019ept}  which computed the SMEFT-LEFT matching and provided the relevant equations in an ancillary {\tt Mathematica} notebook. Renormalization group evolutions (RGE) from the reference scale at which we define the Wilson coefficients $\cb$ and $\cw$, which we take to be the $\mu$C center of mass energy, to the matching scale are included via {\tt DSixTools}\,\cite{Celis:2017hod,Fuentes-Martin:2020zaz}.
As for the theory prediction one has\,\cite{Misiak:2020vlo}
\be
10^4 \times {\cal BR}(B\to X_s \gamma) = (3.39 \pm 0.17) - 2.10\,{\rm Re}[3.93\,C_7 + C_8] \ ,
\ee
at the matching scale $\mu_{\rm EW}=2 m_W$. For the experimental value we use the current world average\,\cite{HeavyFlavorAveragingGroupHFLAV:2024ctg}
\be
10^4 \times  {\cal BR}(B\to X_s \gamma) = 3.49 \pm 0.19 \ .
\ee
The 95\% CL confidence level bounds on $\cb$ and $\cw$ are computed by building a $\chi^2$ distributions summing in quadrature theory and experimental errors and requiring $\Delta\chi^2\le 4$.
For $\cw=0$ and $\cb=0$ one obtains 
$-0.0036\,{\rm TeV}^{-2} \le \cb  \le 0.0050\,$TeV$^{-2}$ and 
$-0.0030\,{\rm TeV}^{-2} \le \cw  \le 0.0042\,$TeV$^{-2}$ respectively fixing the reference scale to $3\,$TeV, in good agreement with the recent analysis of\,\cite{Allwicher:2023shc}\,\footnote{See also\,\cite{Grzadkowski:2008mf,Drobnak:2011wj,Drobnak:2011aa} for earlier related studies.}. The Belle-II experiment aims at reducing the experimental uncertainties from 5.4\% down to 3.2\% with the full dataset of 50\,ab$^{-1}$ of integrated luminosity\,\cite{Belle-II:2018jsg}.

%%%%%%%%%%%%%%%
%%%	MC & STAT	  %%
%%%%%%%%%%%%%%%

\section{$\mu-$collider analysis}\label{sec:results}

At a high-energy $\mu$C the  $\ob$ and $\ow$ operators can be tested via $\mu^+\mu^- \to b \bar b$ scattering, which is sensitive to the dipole structure via the intermediate $s-$channel $\gamma$ or $Z$ bosons.
EW gauge invariance however necessarily implies the presence of interactions with an additional Higgs boson insertion in the dipole structures, see Eq.\,\eqref{eq:ops_EWSB}, allowing to test the dipole operators also via 
other processes, as $\mu^+ \mu^- \to b \bar b h$ scattering. In principle vector boson scattering (VBS) production can offer an additional handle for testing $\ob$ and $\ow$. We however found both neutral- and charged-current VBS scattering processes  to have a reach  comparable to $\mu^+ \mu^- \to b \bar b$ and to be less constraining than $\mu^+\mu^- \to b \bar b h$, in agreement with similar recent analyses of the $\mu$C sensitivity to $t-$quark and $\tau$ dipoles\,\cite{Han:2024gan,Wang:2024bfc,Buttazzo:2026amk}. For this reason we will refrain in presenting a detailed study of VBS processes, restricting to $\mu^+\mu^- \to b \bar b$ and $\mu^+\mu^- \to b \bar b h$ productions only.
\footnote{It is important to note that these processes receive contributions from other $d=6$ SMEFT operators, both in the bottom and in the muon sector, as, {\emph{e.g.}}, the Yukawa type operator $ |H|^2 (\bar q_L^i H d_R^j)$, the Higgs current operators $(H^\dag i \overleftrightarrow D_\mu H )(\bar \ell^i \gamma^\mu \ell^j)$, $(H^\dag i \overleftrightarrow D_\mu^a H )(\bar \ell^i \sigma^a \gamma^\mu \ell^j)$ and the analogous ones with $b-$quark currents and various four-fermion operators. A complete global fit of all $d=6$ deformations affecting the $b\bar b$ and $b\bar b h$ processes at a high-energy \muc~is beyond the scope of the present work. However, we note that projected constraints on these operators at the end of the HL-LHC programme might still allow for sizable  effects in the process under consideration, while angular observables could help disentangle these effects against those of the bottom dipole operators, owing to the different chiral structure. 
}

\subsection{Simulation details and statistical procedure}\label{sec:stat}

To perform our analysis we expand the cross-sections of the relevant processes in terms of their linear and quadratic contributions in the BSM Wilson coefficients $\cb$ and $\cw$. Practically, we parametrize the relative deformation with respect to the SM predictions as
\be\label{eq:expansion}
\frac{\sigma_{\rm NP}}{\sigma_{\rm SM}}=\frac{\sigma_{\cb,\cw} - \sigma_{\rm SM}}{\sigma_{\rm SM}}  = 
\cb\, a_{1,{\cal B}}+
\cb^2\, a_{2,{\cal B}}+
\cw\, a_{1,{\cal W}}+
\cw^2\, a_{2,{\cal W}}+
\cb\,\cw\, a_{2,{\cal B}{\cal W}} \ ,
\ee
where $\sigma_{\rm NP}$, $\sigma_{\cb,\cw}$ and $\sigma_{\rm SM}$ are the NP, total and SM cross-section for the process under consideration respectively. The linear $a_{1,{\cal B}}$ and $a_{1,{\cal W}}$ and the quadratic $a_{2,{\cal B}}$, $a_{2,{\cal W}}$ and $a_{2,{\cal B}{\cal W}}$ coefficients are process, $\sqrt s$ and phase space dependent, and are evaluated through a Monte Carlo simulation performed with {\tt MadGraph5\_aMC@NLO}\,\cite{Alwall:2014hca} for which we have implemented the operators of Eq.\,\eqref{eq:ops} into the {\tt FeynRules}\,\cite{Alloul:2013bka} package using the {\tt UFO}\,\cite{Degrande:2011ua} format.\footnote{With {\tt MadGraph5\_aMC@NLO} it is possible to restrict the interaction order of the various couplings not only at the level of the matrix elements but also at the level of matrix elements squared, allowing to easily extract the linear and quadratic terms in the expansion of Eq.\,\eqref{eq:expansion}.} Our analysis is conducted at the level of a fast detector simulation. Parton showering is performed via {\tt Pythia8}\,\cite{Sjostrand:2014zea} and, when needed, unstable particles are decayed via {\tt MadSpin}\,\cite{Artoisenet:2012st}. We model the detector response via {\tt Delphes3}\,\cite{deFavereau:2013fsa}
using the default {\tt delphes\_card\_MuonColliderDet.tcl} detector card.
Jets are reconstructed via {\tt FastJet}\,\cite{Cacciari:2011ma} using the Valencia clustering algorithm\,\cite{Boronat:2014hva}, which is more suited for an analysis at a lepton colliders than the anti$-\kappa_t$ one. We work in exclusive mode for the number of reconstructed jets and tune the algorithm parameters to $\beta = \gamma = 1$ and $R = 0.5$.  The $b-$jets reconstruction efficiency is fixed to the working point $\epsilon_b=0.75$, conservatively taken from a recent CMS work for boosted jets\,\cite{CMS-DP-2023-065}.

Once the expansion coefficients of Eq.\,\eqref{eq:expansion} are computed and the signal yield in function of $\cb$ and $\cw$ is determined, we evaluate the two dimensional Poisson distribution
\be
p(n_{{\rm obs}}| n_{{\rm th}}) = \frac{1}{n_{{\rm obs}}!}e^{-n_{{\rm th}}}n_{{\rm th}}^{n_{{\rm obs}}} \ ,
\ee
where $n_{\rm th}$ and $n_{\rm obs}$ are the total number of predicted events, function of $\cb$ and $\cw$, and the observed number of events, which for a sensitivity projection we set equal to the SM expectation, and compute 95\% CL intervals in the 
$\cw-\cb$ plane. Our analysis is performed at leading order in the coupling expansion. As is well known EW corrections can become sizable at a high energy \muc~due to the emergence of Sudakov double logarithms that scale as $\frac{\alpha_W}{4\pi}\log^2\frac{s}{m_W^2}$ see, {\emph{e.g.},\,\cite{Accettura:2023ked}. By only rescaling the SM background predictions, these will affect the limits on the $\cb$ and $\cw$ Wilson coefficients of ${\cal O}(5\%)$ for $\sqrt s=10\,$TeV.

As regarding the $\mu$C specifications, throughout our analysis we assume the standard quadratic scaling of the integrated luminosity with respect to the center of mass energy\,\cite{Accettura:2023ked}
\be
{\cal L } =10\,{\rm ab}^{-1}\left( \frac{\sqrt s}{10\,{\rm TeV}}\right)^2 \ 
\ee
which we vary from from 3\,TeV to 14\,TeV.\footnote{Practically, we simulate Monte Carlo events for the standard $\mu-$collider benchmark energies of $\sqrt s=3,\,6,\,10,\,14\,$TeV}

%%%%%%%%%%%%%%%%%%%%%
%%			ANALYSIS 		%%
%%%%%%%%%%%%%%%%%%%%%

\subsection{$\mu^+ \mu^- \to b \bar b$ process}\label{sec:mumu_bb}

\begin{figure}[t!]
\begin{center}
\includegraphics[width=0.32\textwidth]{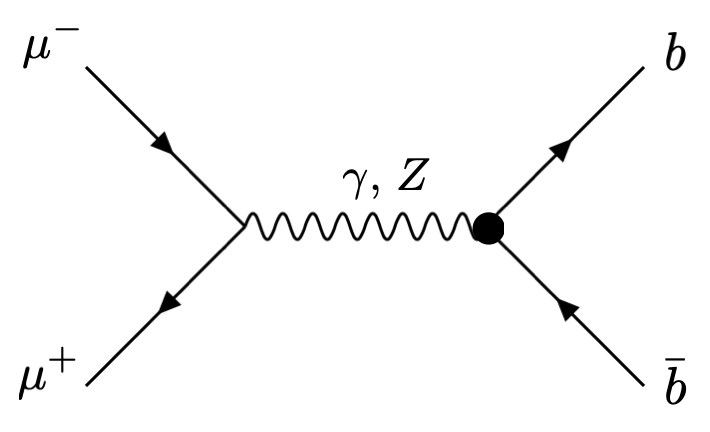}
\end{center}
\caption{
Feynman diagram for the BSM contribution to the process $\mu^+ \mu^- \to b \bar b$  via the $d=6$ dipole operators of Eq.\,\eqref{eq:ops_EWSB}, whose insertions are represented as a black dot.
}
\label{fig:diagrams_bb}
\end{figure}

We start by investigating the $2\to 2$ process $\mu^+ \mu^- \to b \bar b$ which, at leading order, proceeds via a $\gamma/Z$ boson propagating in $s-$channel with modified interactions to the $b-$quark current as shown in Fig.\,\ref{fig:diagrams_bb}. Given the opposite chiral structure of the SM and $d=6$ dipole operators interaction, naive dimensional analysis predicts that the cross-sections for the SM, interference and squared contributions scale as $\sigma_{\rm SM} \simeq \frac{1}{E^2}$, $\sigma_{\rm int} \simeq\frac{v}{\Lambda^2}  \frac{m_b}{E^2} $ and $\sigma_{\rm quad} \simeq  \frac{v^2}{\Lambda^4}$ respectively, where
the NP scale $\Lambda$ is related to the dipole Wilson coefficients by $\Lambda \simeq {\cal C}_{{\cal B},\,{\cal W}}^{-1/2}$,
making the leading interference terms chiral suppressed by the $b-$quark mass.

\begin{figure}[h!]
\begin{center}
\includegraphics[width=0.48\textwidth]{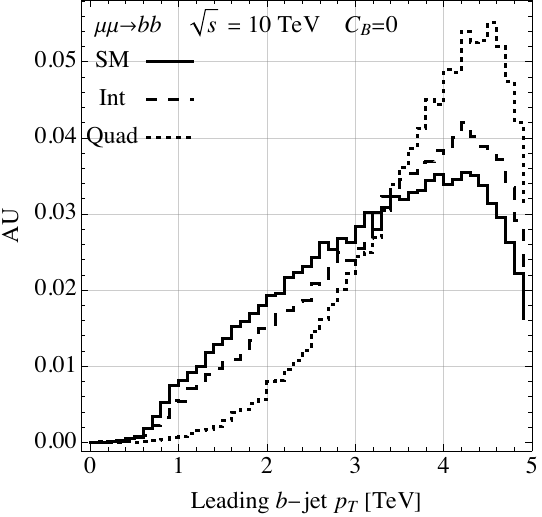}\hfill
\includegraphics[width=0.48\textwidth]{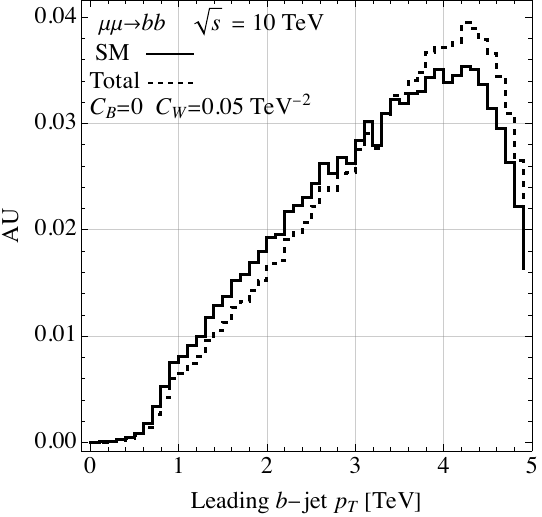}
\hfill
\end{center}
\caption{
Normalised distribution of the leading $b-$jet $p_T$ for the process $\mu^+\mu^-\to b \bar b$ at $\sqrt s= 10\,$TeV. In the left panel we show the SM (solid), interference (dashed) and quadratic (dotted) terms separately, fixing $\cb=0$, while in the right panel we show the SM (solid) and full SM plus $d=6$ contribution (dashed), fixing $\cb=0$ and $\cw = 0.05\,$TeV$^{-2}$.
 }
\label{fig:distr_bb}
\end{figure}

For our analysis we generate parton level events with $p_T^b>20\,$GeV, $|\eta_b|<2.44$ and $\Delta R(b,\bar b)>0.2$.
Following the general strategy described in Sec.\,\ref{sec:stat}, at detector level we reconstruct exactly two $b-$jets with $|\eta^b|<2.44$ via the Valencia algorithm working in exclusive mode. The leading $b-$jet $p_T$ normalised distribution for $\sqrt s = 10\,$TeV and $\cb=0$ is shown in the left panel of Fig.\,\ref{fig:distr_bb}, where the solid line represents the SM, and the dashed and dotted lines the interference and quadratic contribution of the $\ow$ operator respectively. In the right panel we show instead the SM and the full SM plus $d=6$ contribution for the representative choice $\cb = 0$ and $\cw = 0.05\,$TeV$^{-2}$, again for $\sqrt s=10\,$TeV.
We see that the  $d=6$ dipole operator produces a harder transverse momentum for the leading $b-$jet, with the second $b-$jet presenting a  similar behavior. We then maximise the reach in the $\cw$$-$$\cb$ plane, by varying the $b-$jet  $p_T$ threshold, obtaining and optimal cut of $p_T^{b}\gtrsim0.15\,\sqrt s\,$.
We  report in Tab.\,\ref{tab:coeff_bb} the expansion coefficients of Eq.\,\eqref{eq:expansion}  for some representative $p_T^b$ cuts for $\sqrt s =3,\,10\,$TeV.\footnote{Note that the $b-$jet reconstruction efficiency $\epsilon_b$ factors out in the ratio with the SM cross-section.} From the table we see that the linear coefficients $a_{1,{\cal B}}$ and $a_{1,{\cal W}}$ are almost constant with the increase of $\sqrt s$, while the quadratic coefficients $a_{2,{\cal B}}$, $a_{2,{\cal W}}$ and $a_{2,{\cal B}{\cal W}}$ present an approximate quadratic scaling with the $\mu$C center of mass energy, in accord with the naive intuition derived from the interaction chiral structure.

The 95\% CL single parameter and marginalised limits on the $\cw$ and $\cb$ Wilson coefficients are reported in Tab.\,\ref{tab:proj} for $\sqrt s =3,\,10\,$TeV, while the projected sensitivities in the $\cw$$-$$\cb$ plane are shown in Fig.\,\ref{fig:exclusion} for the same $\sqrt s$ values. There we also overlay the projected limit from the $R_b$ measurement at FCC-ee/CEPC (which are approximately equal) and the one arising from $B\to X_s\gamma$. From the figures we see that the study of the $\mu^+ \mu^- \to b \bar b$ scattering at a 3\,TeV $\mu$C will greatly improve upon the $R_b$ limits set by LEP and SLD discussed in Sec.\,\ref{sec:Rb}, which are not shown in the plot, offering a comparable sensitivity to the projected $R_b$ limit from CEPC, also closing the ${\cal C}_Z=0$ flat direction. The constraints we obtain from this process are stronger that the one that will be obtained at the end of the HL-LHC, see {\emph{e.g.}},\,\cite{Durieux:2019rbz}. Interestingly, at $\sqrt s = 10\,$TeV, the analysis of this process starts approaching the indirect limits arising from the $\Delta F=1$ transition. Also note that for this process the effective NP scale $\Lambda$ that can be tested lies at the edge of the validity of the effective descriptions for the $\ob$ and $\ow$ operators, being close to the chosen center of mass energy of the $\mu$C. These aspects are clearly visible in Fig.\,\ref{fig:exclusion_energy}, where we show the single parameter limits on the positive values of ${\cal C}_{\cal B}$ and ${\cal C}_{\cal W}$ in function of $\sqrt s$. There the gray shaded area represents the region where $\sqrt s > {\cal C}_{{\cal B},\,{\cal W}}^{-1/2}$ and the EFT cannot be trusted, while the solid and dashed horizontal green lines are the current and projected limits on the $B \to X_s\gamma$ rate.

\begin{table}
\begin{center}
\scalebox{0.80}{
\begin{tabular}{c|c|cccccc}
\multicolumn{8}{c}{{\bf{Expansion coefficients for $\mu\mu \to bb$}}}
\vspace{0.2cm}  \\
 $\sqrt{s}\,$[TeV]  & 
 $p_{T,{\rm min}}\,$[GeV]& 
 $\sigma_{\rm SM}\,$[fb] 
 &  $a_{1,{\cal B}}\,$[TeV$^2$] 
 &  $a_{2,{\cal B}}\,$[TeV$^4$] 
 &  $a_{1,{\cal W}}\,$[TeV$^2$] 
 &  $a_{2,{\cal W}}\,$[TeV$^4$] 
 & $a_{2,{\cal BW}}\,$[TeV$^4$] \\
 \hline
 \multirow{2}{*}{3} 
 & 500
 &   $7.36$ 
 &   $-0.025$
 &   $18.3$
 &   $0.034$
 &   $12.0$
 &   $-13.3$	\\
 
 & 1000
 &   $2.44$ 
 &   $-0.028$
 &   $23.3$
 &   $0.04$
 &   $15.3$
 &   $-16.9$	\\
			    \hline
 \multirow{2}{*}{10} 
 & 1500
 &   $0.72$ 
 &   $-0.025$
 &   $195.1$
 &   $0.033$
 &   $128.0$
 &   $-140.9$	\\
 
 & 3000
 &   $0.31$ 
 &   $-0.027$
 &   $248.9$
 &   $0.037$
 &   $166.1$
 &   $-182.4$	\\
			    \hline			    
\end{tabular}
}
\end{center}
\caption{Coefficients of the expansion of Eq.\,\eqref{eq:expansion} for the $\mu\mu\to bb$ process.}
\label{tab:coeff_bb}
\end{table}

\subsection{$\mu^+ \mu^- \to b \bar b h$ process}\label{sec:mumu_bbh}

\begin{figure}[h!]
\begin{center}
\includegraphics[width=0.32\textwidth]{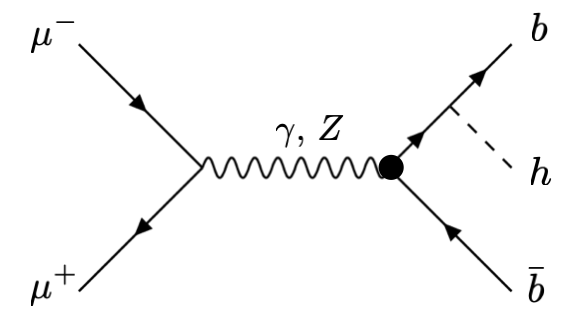}\hfill
\includegraphics[width=0.32\textwidth]{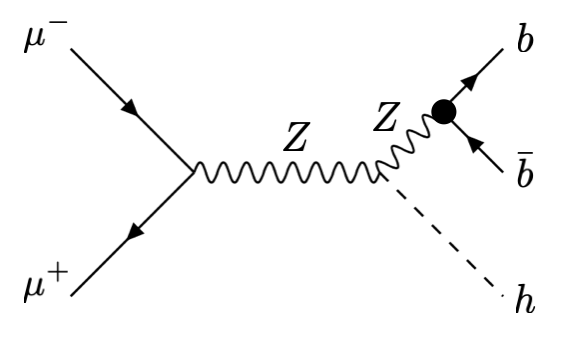}\hfill
\includegraphics[width=0.32\textwidth]{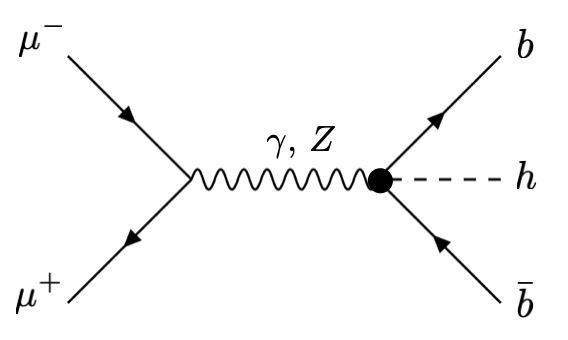}
\end{center}
\caption{
Feynman diagrams for the BSM contribution to the process $\mu^+ \mu^- \to b \bar b h$  via the $d=6$ dipole operators of Eq.\,\eqref{eq:ops_EWSB}, whose insertions are represented as a black dot.
}
\label{fig:diagrams_bbh}
\end{figure}

\begin{figure}[h!]
\begin{center}
\includegraphics[width=0.48\textwidth]{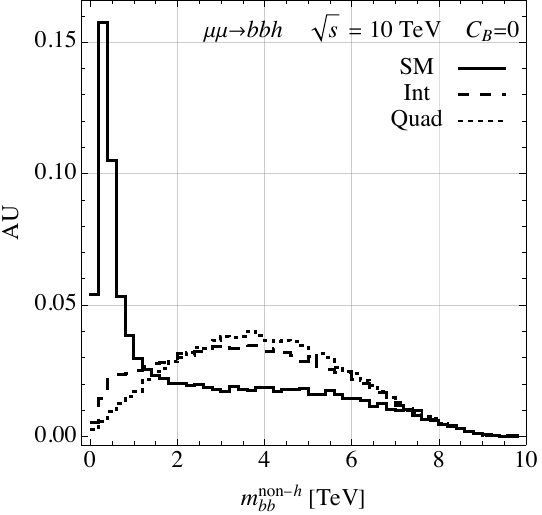}\hfill
\includegraphics[width=0.48\textwidth]{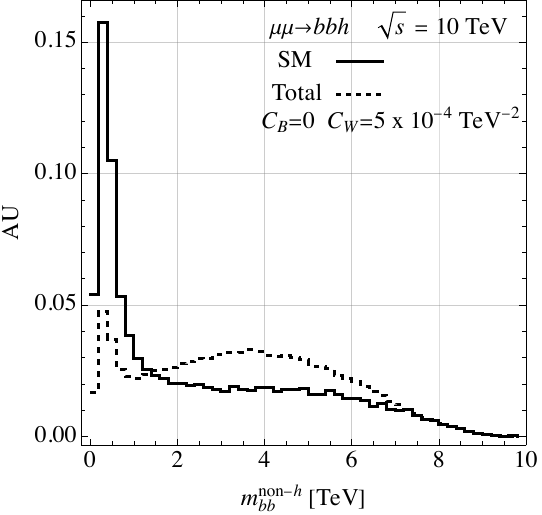}
\hfill
\end{center}
\caption{
Normalised distribution of the invariant mass of the two $b-$jets not reconstructing the Higgs mass for the process $\mu^+\mu^-\to b \bar b h $ at $\sqrt s= 10\,$TeV. In the left panel we show the SM (solid), interference (dashed) and quadratic (dotted) terms separately, fixing $\cb=0$, while in the right panel we show the SM (solid) and full SM plus $d=6$ contribution (dashed), fixing $\cb=0$ and $\cw = 5\times 10^{-4}\,$TeV$^{-2}$. }
\label{fig:distr_bbh}
\end{figure}

We then move to the $2\to 3$ process $\mu^+ \mu^- \to b \bar b h$, whose BSM contribution proceeds via the diagrams of Fig.\,\ref{fig:diagrams_bbh}. To maximise the rate we will consider the Higgs boson decaying into a pair of $b-$quarks, with a rate BR($h\to b \bar b)=58.24\%$. We anticipate that for our analysis we will select events asking for a large invariant mass for the $b-$jets not reconstructing the Higgs boson, thereby effectively removing the contribution from the Higgs-strahlung diagram represented in the central panel. Hence the SM contribution only proceeds via the left-panel diagram, which has a chiral flip due to the Higgs boson insertion.
The resulting chiral structure is the same as the one of the BSM contribution given by the Feynman diagram on the right panel, where the chiral flip is given by the insertion of the dipole operator. Hence the SM and BSM amplitude can have the same chiral structure.
 Naive dimensional analysis predicts that the cross-sections for the SM, interference and squared contributions now scale as $\sigma_{\rm SM} \simeq \frac{1}{E^2}$, $\sigma_{\rm int} \simeq \frac{1}{\Lambda^2}$ and $\sigma_{\rm quad} \simeq  \frac{E^2}{\Lambda^4}$ respectively, lifting the chiral suppression for the interference term that is present in the $\mu^+\mu^- \to b \bar b$ process.
  
 For our analysis we now generate parton level events for the process $\mu^+\mu^- \to b \bar b h$ with $p_T^b>100\,$GeV, $|\eta^b|<2.44$ and $\Delta R(b,\bar b)>0.2$ and decay the Higgs boson via {\tt MadSpin}. Following again the general strategy described in Sec.\,\ref{sec:stat} at detector level we reconstruct exactly four $b-$jets with $p_T^b > 100\,$GeV and $|\eta^b|<2.44$ via the Valencia algorithm working in exclusive mode. We then reconstruct the Higgs candidate with the following procedure. We compute the invariant masses of for all the $b-$jet pairs and identify as Higgs candidate the one having the minimum angular separation $\Delta R$ with the requirement $|m_{bb}-m_h|<25\,$GeV. If no $b-$jet pair is present within this window, we identify as Higgs candidate the one minimising $|m_{bb}-m_h|$. 
Still following\,\cite{CMS-DP-2023-065}, we then also apply a reconstruction efficiency $\epsilon_h=0.75$ on the Higgs jet. We then show in Fig.\,\ref{fig:distr_bbh} the invariant mass of the two $b-$jets not reconstructing the Higgs mass. In the left panel we fix $\sqrt s = 10\,$TeV and $\cb=0$ and  show the SM, interference and quadratic contributions from the $\ow$ operator, while in the right panel we fix $\sqrt s = 10\,$TeV, $\cb=0$ and $\cw = 5\times 10^{-4}\,$TeV$^{-2}$ and show the SM and full SM plus $d=6$ contribution. We see that the  $d=6$ dipole operator produces a harder invariant mass for the two $b-$jets. We then maximise the reach in the $\cw$$-$$\cb$ plane, by varying the $m_{bb}$ threshold, obtaining and optimal cut of $m_{bb}\gtrsim0.06\,\sqrt s\,$.

\begin{table}
\begin{center}
\scalebox{0.80}{
\begin{tabular}{c|c|cccccc}
\multicolumn{8}{c}{{\bf{Expansion coefficients for $\mu\mu \to bbh$}}}
\vspace{0.2cm} \\
 $\sqrt{s}\,\,$[TeV]  & 
 $m^{\rm min}_{bb}\,$[GeV]& 
 $\sigma_{\rm SM}\,$[fb] 
 &  $a_{1,{\cal B}}\,$[TeV$^2$] 
 &  $a_{2,{\cal B}}\,$[TeV$^4$] 
 &  $a_{1,{\cal W}}\,$[TeV$^2$] 
 &  $a_{2,{\cal W}}\,$[TeV$^4$] 
 & $a_{2,{\cal BW}}\,$[TeV$^4$] \\
 \hline
 \multirow{2}{*}{3} 
 & 200
 &   $3.8\times 10^{-4}$ 
 &   $-67.7$
 &   $6.78\times 10^4$
 &   $91.9$
 &   $4.46 \times 10^4$
 &   $-4.93\times 10^4$	\\
 
 & 400
 &   $1.64\times 10^{-4}$ 
 &   $-144.6$
 &   $1.46\times 10^5$
 &   $195.7$
 &   $9.58 \times 10^4$
 &   $-1.05\times 10^5$	\\
			    \hline
 \multirow{2}{*}{10} 
 & 600
 &   $1.99\times 10^{-5}$ 
 &   $-1.70\times 10^{3}$
 &   $1.85\times 10^7$
 &   $2.27\times 10^3$
 &   $1.21 \times 10^7$
 &   $-1.34\times 10^7$	\\
 
 & 1200
 &   $1.40\times 10^{-5}$ 
 &   $-2.25\times 10^{3}$
 &   $2.49\times 10^7$
 &   $3.01\times 10^3$
 &   $1.64 \times 10^7$
 &   $-1.81\times 10^7$	\\
			    \hline			    
\end{tabular}
}
\end{center}
\caption{Coefficients of the expansion of Eq.\,\eqref{eq:expansion} for the $\mu\mu\to bbh$ process.}
\label{tab:coeff_bbh}
\end{table}

We report in Tab.\,\ref{tab:coeff_bbh} the expansion coefficients of Eq.\,\eqref{eq:expansion}  for some representative $m_{bb}$ cuts.\footnote{Again note that the $b-$jet and Higgs reconstruction efficiencies $\epsilon_b$ and $\epsilon_h$, together with the branching ratio for $h\to bb$, factor out in the ratio with the SM cross-section.} Here we see that the linear $a_{1,{\cal B}}$ and $a_{1,{\cal W}}$ and quadratic $a_{2,{\cal B}}$, $a_{2,{\cal W}}$ and $a_{2,{\cal B}{\cal W}}$ coefficients present an approximate quadratic and quartic scaling with the $\mu$C center of mass energy respectively, in accord with the naive intuition derived from the interaction chiral structure.

The 95\% CL single parameter and marginalised limits on the $\cw$ and $\cb$ Wilson coefficients are reported in Tab.\,\ref{tab:proj} while the projected sensitivities in the $\cw$$-$$\cb$ plane are shown in Fig.\,\ref{fig:exclusion} for both $\sqrt s =3\,$TeV and $\sqrt s =10\,$TeV together with the projected limits from the $\mu^+ \mu^- \to b\bar b$ process and the indirect limits from EW and flavor precision measurements. From the figures we see that the study of the $b\bar b h$ process allows to set a comparable, but stronger, limit with respect to the $b\bar b$ one, already offering a better sensitivity than the one attainable from the measurement from $R_b$ at FCC-ee/CEPC already at $\sqrt s = 3\,$TeV. The benefit of a higher $\sqrt s$ is evident from the right panel of Fig.\,\ref{fig:exclusion}. At $\sqrt s=10\,$TeV the limits from the $b\bar b h$ channel are much greater than that from the $bb$ one, reaching a sensitivity comparable to the ones derived by the measurements of the $\Delta F=1$ transition $B\to X_s\gamma$ and allowing to completely close one if its flat direction.
For this process the effective NP scale $\Lambda$ that can be tested is well within the validity of the effective descriptions for the $\ob$ and $\ow$ operators, being larger than the chosen center of mass energy of the $\mu$C, which can be again seen in Fig.\,\ref{fig:exclusion_energy}, where we show the single parameter limits on the positive values of ${\cal C}_{\cal B}$ and ${\cal C}_{\cal W}$ in function of $\sqrt s$, together with the region where the EFT description ceases to be valid and the current and projected limits on the $B \to X_s\gamma$ rate.

\begin{table}[t!]
\begin{center}
\scalebox{0.80}{
\begin{tabular}{c|c||c|c||c|c}
\multicolumn{2}{c}{} & \multicolumn{2}{c}{${\cal C}_{\cal B}\,[{\rm TeV}^{-2}]\times 10^{-2}$} & \multicolumn{2}{c}{${\cal C}_{\cal W}\,[{\rm TeV}^{-2}]\times 10^{-2}$} \\
 $\sqrt{s}\,\,$[TeV]  
 &  Process
 & Single parameter
 &  Marginalised
 & Single parameter
 &  Marginalised\\
 \hline
 \multirow{2}{*}{3} 
 &  $\mu^+ \mu^- \to b \bar b$
 &   $[-4.1,\,+4.4]$ 
 &   $[-4.7,\,+4.7]$ 
 &   $[-5.4,\,+5.0]$  
 &   $[-6.0,\,+5.7]$ 	\\
 &  $\mu^+ \mu^- \to b \bar b h$
 &   $[-2.3,\,+2.4]$  
 &   $[-2.6,\,+2.7]$ 
 &   $[-3.0,\,+2.8]$  
 &   $[-3.3,\,+3.1]$ 	\\
			    \hline
 \multirow{2}{*}{10} 
 &  $\mu^+ \mu^- \to b \bar b$
 &   $[-1.3,\,+1.2]$  
 &   $[-1.4,\,+1.4]$ 
 &   $[-1.6,\,+1.4]$ 
 &   $[-1.8,\,+1.7]$ 	\\
 &  $\mu^+ \mu^- \to b \bar b h$
 &   $[-0.18,\,+0.19]$  
 &   $[-0.20,\,+0.21]$ 
 &   $[-0.24,\,+0.20]$ 
 &   $[-0.27,\,+0.24]$ 	\\
			    \hline			    
\end{tabular}
}
\end{center}
\caption{Single parameter and marginalised limits at 95\% CL on the $\cb$ and $\cw$ Wilson coefficients from the $\mu^+ \mu^- \to b \bar b$ and  $\mu^+ \mu^- \to b \bar b h$ processes.}
\label{tab:proj}
\end{table}

\begin{figure}[t!]
\begin{center}
\includegraphics[width=0.48\textwidth]{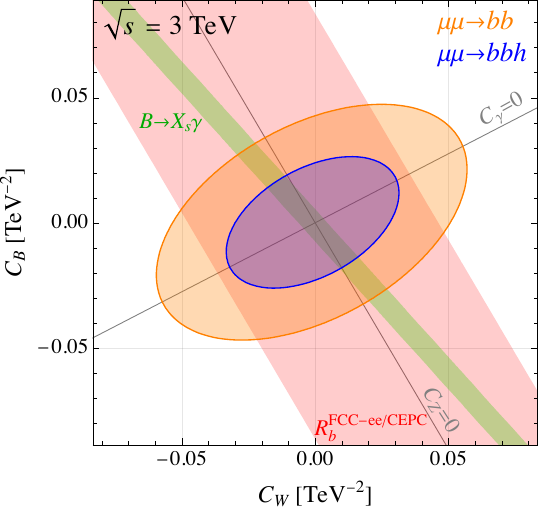}\hfill
\includegraphics[width=0.49\textwidth]{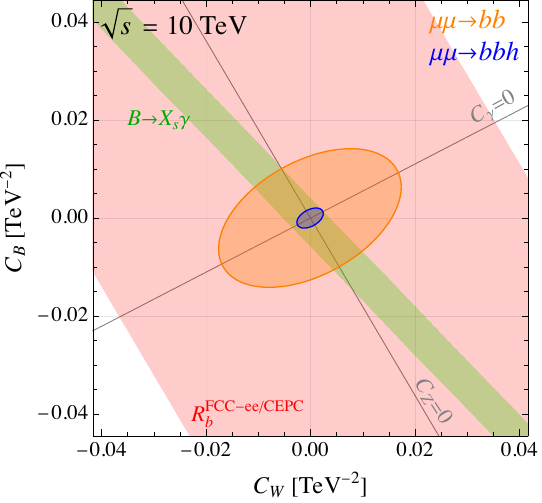}
\hfill
\end{center}
\caption{
Exclusion limits at 95\% CL on the $\ow$ and $\ob$ operators in the $\cw-\cb$ plane for a $\mu$C with $\sqrt s=3\,$TeV (left) and $\sqrt s=10\,$TeV (right) from the process $\mu^+\mu^- \to b\bar b$ (orange) and $\mu^+\mu^- \to b\bar bh$ (blue). Also shown are the current limits from the $\Delta F=1$ transition $B \to X_s \gamma$ and the projected limits from the $R_b$ measurement at CEPC. The gray lines indicate the directions where the ${\cal C}_\gamma$ and ${\cal C}_Z$  coefficients of Eq.\,\eqref{eq:cgamam_cz} vanish.
}
\label{fig:exclusion}
\end{figure}

\begin{figure}[t!]
\begin{center}
\includegraphics[width=0.49\textwidth]{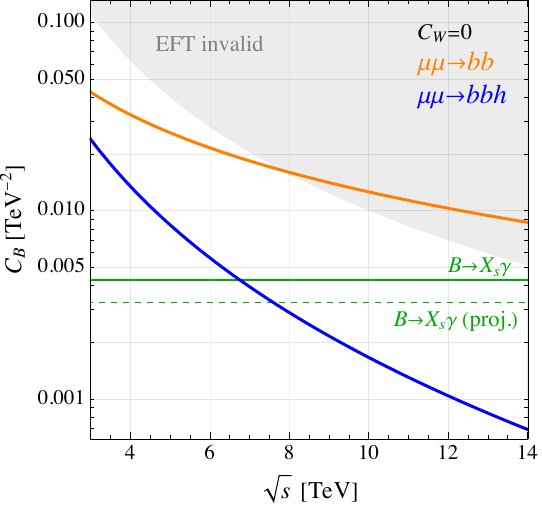}\hfill
\includegraphics[width=0.49\textwidth]{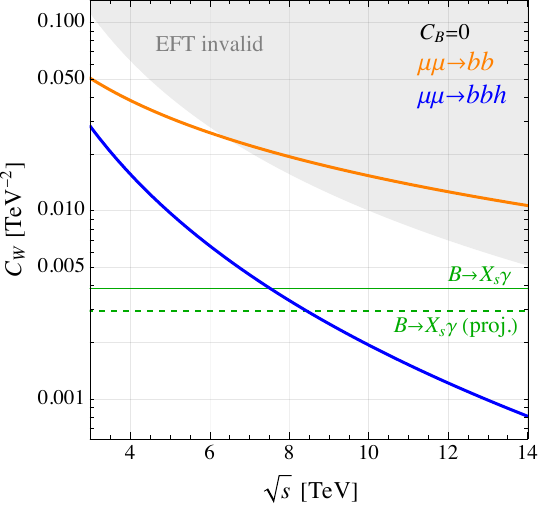}
\hfill
\end{center}
\caption{
Single parameter limits at 95\% CL on the positive $\cb$ (left) and $\cw$ (right) Wilson coefficients in function of the $\mu-$collider center of mass energy.  Also shown are the current bounds (solid green line) and projected limits (dashed green line) from the  $\Delta F=1$ transition $B \to X_s \gamma$. The gray shaded area indicates the region where 
$\sqrt s > {\cal C}_{{\cal B},\,{\cal W}}^{-1/2}$ and the EFT cannot be trusted.}
\label{fig:exclusion_energy}
\end{figure}

\section{Conclusions}\label{sec:concl}

Assessing the potential of future colliders in testing possible deviations from the SM predictions is an important goal of current research in high-energy particle physics. In this work, we have investigated the sensitivity of a multi--TeV $\mu-$collider in probing EW dipole moments of the bottom quark. We have parametrized the relevant deformation of the SM Lagrangian in the language of the SMEFT and reviewed current limits and future prospects on the $d=6$ operators $\ob$ and $\ow$ arising from EW and flavor precision measurements. Then, working at the level of a fast detector simulation, we have studied the processes $\mu^+\mu^- \to b\bar b$ and $\mu^+\mu^- \to b\bar b h$. In particular we have shown that the latter offers a better sensitivity with respect to the former, owing to the specific chiral structure of the dipole interaction. At $\sqrt  s=3\,$TeV, one obtains  sensitivities comparable to the projected ones of FCC-ee/CEPC  arising from the measurements of $R_b$. For higher $\sqrt s$ it is possible to improve on current and projected limits obtainable from the measurement of the
$\Delta F=1$ transition $B\to X_s\gamma$, while remaining within the EFT validity of the $\ow$ and $\ob$ dipole operators when interpreted in terms of a strongly coupled UV completion.

\section*{Acknowledgments}

We thank Lukas Allwicher for useful discussions.

%%%%%%%%%%%%%%%%%
%%%	REFERENCES	   %%%		
%%%%%%%%%%%%%%%%%

\bibliographystyle{JHEP}
{\footnotesize
\bibliography{biblio}}

\end{document}